\documentclass[a4paper,12pt]{article}

\newcommand{\majhn}{\lambda}
\newcommand{\ndl}{{\bf B}}
\newcommand{\tir}{\vec{r}}
\newcommand{\hit}{\vec{v}}
\newcommand{\htr}{{\bf \beta}}
\newcommand{\ohtr}[1]{\htr^{(#1)}}
\newcommand{\sila}{F}
\newcommand{\Fext}{F_{\rm ext}}
\newcommand{\fext}{{\bf f}}

\title{Supplementary kinetic constants of charged particles}
\author{Marijan Ribari\v c and Luka \v Su\v ster\v si\v c\thanks{Corresponding author. Phone +386 1 477 3258; fax +386 1 423 1569; electronic address: \tt luka.sustersic@ijs.si\rm} \\Jo\v zef Stefan Institute, p.p.3000, 1001 Ljubljana, Slovenia}
\date{}

\begin{document}

\maketitle

\begin{abstract}

We put forward: (A)~An improved description of classical, kinetic properties of a charged pointlike physical particle that consists, in addition to its mass and charge, also of the Eliezer and Bhabha kinetic constants; and (B)~a proposal to evaluate these kinetic constants by considering the trajectories of charged particles in an acccelerator.

\end{abstract}
\vskip 0.5 in
\noindent PACS numbers: 45.50.Dd; 06.30.Gv; 29.27.Bd

\noindent Keywords: Charged particle trajectories; kinetic constants; accelerator physics
\vfill\eject

We consider classical, kinetic properties of a charged physical particle with mass and charge concentrated around a point $\tir(t)$ moving with velocity $\hit(t)$ under the influence of an external force $\Fext(r,t)$. Ignoring its size, let us approximate its behaviour by the following model.

\it Charged point particle. \rm (i)~The charged point particle with mass $m$ and charge $q$ located at $\tir(t)$ moves under the influence of the external force $\Fext(\tir(t),t)$. (ii)~The relation between the dimensionless four-force
\begin{equation}
   \fext(t) \equiv {\tau_0\over mc}\, \gamma(t) \Bigl( \hit(t)\cdot \Fext(\tir(t),t) /c, 
            \Fext(\tir(t),t) \Bigr) \,,
   \label{fourforce}
\end{equation}
with $\gamma(t) \equiv \sqrt{1 - |\hit(t)/c|^2}$ and $\tau_0 \equiv q^2/6\pi\epsilon_0 mc^3$, and the dimensionless four-velocity $\htr(t) \equiv (\gamma, \gamma\hit/c)$ is \it invariant under Poincar\'e transformations; \rm (iii)~The charged point particle emits four-momentum through its Lienard-Wiechert potentials; and (iv)~The relativistic differential energy-momentum balance equation for the charged point particle reads
\begin{equation}
   \ohtr{1} - \bigl( \ohtr{1} \cdot \ohtr{1} \bigr) \htr + \ndl^{(1)} = \fext \,,
   \qquad {}^{(n)} \equiv (\tau_0 \gamma d/dt)^n \,,
   \label{energycons}
\end{equation}
where, generalizing Schott \cite{Schott}, we introduced the dimensionless acceleration four-momentum $\ndl(t)$; we use the metric with signature $(+---)$, so that $\htr\cdot\htr =1$. Now, Dirac \cite{Dirac} concluded that the conservation of four-momentum (\ref{energycons}) requires that
\begin{equation}
   \htr \cdot (\ndl + \ohtr1)^{(1)} = 0 \,,
   \label{Diraccond}
\end{equation}
whereas Bhabha \cite{Bhabha} pointed out that conservation of angular four-momentum requires that the cross product
\begin{equation}
   \htr \wedge (\ndl + \ohtr1 )
   \label{Bhabhacond}
\end{equation}
is a total differential with respect to proper time, cf.~\cite{mi003}, Part~II. These two conditions point out that $\ndl$ is an essential part of the balance equation (\ref{energycons}).

Relation (\ref{energycons}) could be used as a Newtonian equation of motion for a charged point particle, were $\ndl$ given only as a function of velocity nad external force. So far no such case is known, cf.~\cite{mi001}. In particular, no Newtonian equation of motion is known for an electron despite an ongoing, century-old quest.

\it Asymptotic differential relation for trajectories of a charged point particle. \rm We put forward arguments in support of the following hypothesis \cite{mi002, mi003}: When the external force depends on a non-negative parameter $\majhn$ in such a way that $\Fext(t,\majhn) = \majhn\sila(\majhn t)$, $\sila(t)$ being an analytic function of $t > 0$ and $\sila(t\le0) = 0$, then in the asymptote $t \to \infty$ the $n$th derivative $d^n\hit(t)/dt^n$ of the particle velocity is of the order $\majhn^n$ as $\majhn \to 0$, and we may approximate the acceleration four-momentum $\ndl$ up to the order of $\majhn^5$ inclusive so that the trajectory of a charged point particle satisfies in the asymptote $t \to \infty$ up to the order of $\majhn^6$ inclusive the following relativistic differential relation:
\begin{eqnarray}
   &&\ohtr{1} - \bigl( \ohtr{1} \cdot \ohtr{1} \bigr) \htr - \ohtr{2} \nonumber\\
   &&\qquad + e_1 \Bigl[ \ohtr{2} - \bigl( \htr\cdot\ohtr{2} - 
      {\textstyle{1\over2}} \ohtr{1}\cdot\ohtr{1} \bigr)\htr \Bigr] ^{(1)} \label{relacija} \\
   &&\qquad + e_2 \Bigl[ \ohtr4 - \bigl( \htr\cdot\ohtr4 - \ohtr1 \cdot\ohtr3 +
      {\textstyle{1\over2}} \ohtr2 \cdot\ohtr2 \bigr)\htr \Bigr] ^{(1)}  \nonumber\\
   &&\qquad + b_1 \Bigl[ \bigl( \ohtr1 \cdot\ohtr1 \bigr)\ohtr2 +2 \bigl(\ohtr1 \cdot\ohtr2
      \bigr)\ohtr1 + {\textstyle{7\over4}} \bigl(\ohtr1 \cdot\ohtr1 \bigr)^2 \htr \Bigr] ^{(1)}
      = \fext \,, \nonumber
\end{eqnarray}
where $e_1$, $e_2$ and $b_1$ are real parameters. Relativistic polynomials in $\ohtr{n}$ multiplied by $e_1$ and $e_2$ were constructed by Eliezer \cite{Eliezer}, and that multiplied by $b_1$ is due to Bhabha \cite{Bhabha}. So let us refer to $e_1$ and $e_2$ as the Eliezer constants, and to $b_1$ as the Bhabha constant.

\it Evaluation of Eliezer and Bhabha constants. \rm So far, the Eliezer and Bhabha constants in (\ref{relacija}) are just parts of a model, the charged point particle. Whether we can use them to describe classical, kinetic properties of a particular charged physical particle, supplementary to its mass and charge, could be determined by observing accelerated trajectories of this particle. To this end we could use particle accelerators, since they may accelerate particles of any mass, from electrons and positrons to uranium ions: a potential use of accelerators for considering also classical, kinetic properties of physical particles.

Considering the kinetic properties of an electron (or positron), one might check the Dirac assumption \cite{Dirac} that an electron is such a simple thing that $\ndl = \ohtr1 $; so that the Eliezer and Bhabha constants of an electron should turn out to be negligible.


\begin{thebibliography}{99}

\bibitem{Schott}G.A. Schott, Ann. Phys. (Leipzig) \bf25 \rm (1908) 63.

\bibitem{Dirac}P.A.M. Dirac, Proc. Roy. Soc. (London) \bf  A 167 \rm (1938) 148.

\bibitem{Bhabha}H.J. Bhabha, Proc. Indian Acad. Sci. \bf A 10 \rm (1939) 115.

\bibitem{mi001}M. Ribari\v c and L. \v Su\v ster\v si\v c, physics/0511033.

\bibitem{mi002}M. Ribari\v c and L. \v Su\v ster\v si\v c, Phys. Lett. A\bf139 \rm(1989) 5.

\bibitem{mi003}M. Ribari\v c and L. \v Su\v ster\v si\v c, \it Conservation Laws and Open Questions of Classical Electrodynamics, \rm World Scientific, Singapore 1990, Chapters 9--11.

\bibitem{Eliezer}C.J. Eliezer, Proc. Roy. Soc. (London) \bf A 194 \rm (1948) 543.

\end{thebibliography}
\end{document}